\documentclass[%
 reprint,
 superscriptaddress,
 floatfix,
 amsmath, amssymb,
 aps,
 prx,
]{revtex4-2}

\usepackage{enumerate}
\usepackage{graphicx}
\usepackage{dcolumn}
\usepackage{bm}

\usepackage{amsthm}

\usepackage{dsfont}
\usepackage{comment}
\usepackage{xcolor}
\usepackage{braket}
\usepackage[colorlinks=true,linkcolor=blue,citecolor=blue,urlcolor=blue]{hyperref}

\begin{document}

\preprint{APS/123-QED}

\title{Certification of quantum properties with imperfect measurements}

\author{Leonardo Zambrano}
\email{leonardo.zambrano@icfo.eu}
\affiliation{ICFO - Institut de Ciencies Fotoniques, The Barcelona Institute of Science and Technology, 08860 Castelldefels, Barcelona, Spain}

\author{Teodor Parella-Dilmé}
\affiliation{ICFO - Institut de Ciencies Fotoniques, The Barcelona Institute of Science and Technology, 08860 Castelldefels, Barcelona, Spain}

\author{Antonio Ac\'{\i}n}
\affiliation{ICFO - Institut de Ciencies Fotoniques, The Barcelona Institute of Science and Technology, 08860 Castelldefels, Barcelona, Spain}
\affiliation{ICREA - Instituci\'{o} Catalana de Recerca i Estudis Avan\c{c}ats, 08010 Barcelona, Spain.}

\author{Donato Farina}
\affiliation{Physics Department E. Pancini - Università degli Studi di Napoli Federico II, Complesso Universitario Monte S. Angelo - Via Cintia - I-80126 Napoli, Italy}

\date{\today}
\begin{abstract}
The accurate characterization of quantum systems is essential for the advancement of quantum technologies. In particular, certifying convex functions of quantum states plays a central role in many applications. We present a certification method for experimentally prepared quantum states that accounts for both shot noise and measurement imperfections in the data-acquisition stage. Building upon previous work, our method extends confidence regions to accommodate imperfect control over measurements. The values of the functions can then be bounded using convex optimization techniques. We provide explicit prescriptions for quantifying the noise contribution from finite statistics and for estimating the effect of measurement imperfections. By jointly incorporating statistical and systematic errors, the method yields a robust certification framework for quantum experiments.
\end{abstract}

\maketitle

\section{Introduction}

In the rapidly advancing field of quantum information processing, the precise characterization of quantum systems poses a fundamental challenge with far-reaching implications for the development of quantum technologies. In particular, certifying values of convex functions of quantum states is a critical task, since several important quantities, including expectation values of Hamiltonians, quantum fidelities, entropies, and quantum Fisher information, belong to this class.  

Some simple quantities can be directly estimated by measuring the quantum system with specifically designed observables \cite{horodecki1996separability, lewenstein2000optimization, saunders2010experimental, wittmann2012loophole}. For more complex functions, the standard approach is to measure a set of observables to reconstruct the state using quantum tomography, and then apply the function to the estimated state \cite{hradil1997quantum, thew2002qudit, gross2010quantum}. However, these methods often fail to certify the value of the function because they neglect realistic imperfections such as shot noise and measurement errors. Ignoring these effects can lead to large uncertainties that compromise the robustness and accuracy of quantum computations, or result in incorrect conclusions about experiments.

The certification of quantum properties in noisy setups has attracted increasing interest over the last few years \cite{christandl2012reliable, guta2020fast, de2023user, zambrano2024certification, aaronson2018shadow, huang2020predicting, chen2021robust, koh2022classical, morelli2022entanglement, cao2023genuine, tavakoli2024quantum, mortimer2026bounding}. Refs.~\cite{christandl2012reliable, guta2020fast, de2023user, zambrano2024certification} address the effect of shot noise on the estimation of convex functions of quantum states by providing confidence regions for quantum state tomography. In other words, they certify that the experimental quantum state lies within a region of the state space with arbitrary high probability. These confidence regions can then be integrated into convex optimization procedures to obtain certified bounds for convex functions. Since the size of the confidence regions depends on shot noise, the bounds also incorporate this effect, making the estimation more rigorous. All these methods assume that the measurement devices are perfectly characterized, and therefore do not account for possible calibration errors or imperfections in the measurement process.

For many-qubit systems, classical shadow techniques provide a way to estimate functions of quantum states while including confidence intervals that account for shot noise \cite{aaronson2018shadow, huang2020predicting}. However, these methods originally assume ideal projective measurements in the computational basis and thus neglect measurement imperfections. Extensions of the classical shadows approach have been proposed \cite{chen2021robust, koh2022classical}, but these corrections are not fully general since they focus on specific noise models. On the other hand, Refs.~\cite{morelli2022entanglement, cao2023genuine, tavakoli2024quantum} incorporate imperfect measurements, but only in restricted scenarios or protocols in quantum information theory.  

Here, we propose a certification method that accounts for both shot noise and measurement imperfections in quantum systems. Our method extends the confidence-region framework introduced in Ref.~\cite{zambrano2024certification} to include situations where experimenters have imperfect control over measurements. We show how to compute an expanded confidence region that incorporates the discrepancy between the actual measurement and the intended one. The parameters defining the confidence region can be obtained from a prior characterization of the system, either through numerical simulations of the system or through experimental measurements. Once this region is established, we use convex optimization techniques to derive rigorous bounds for convex functions of quantum states. Importantly, the method does not require the measurements to be informationally complete, which reduces the experimental effort needed to apply the protocol. By integrating both shot noise and measurement noise into a single framework, our approach provides a versatile and robust certification tool for quantum systems. 

This article is organized as follows: in Sec.~\ref{sec:prelim} we review the mathematical preliminaries needed for our work. In Sec.~\ref{sec:results} we present our method, discuss its applicability, and show how to calculate the parameters that define the confidence region. In Sec.~\ref{sec:numerical} we test the applicability of the protocol through numerical examples, to finally conclude in Sec.~\ref{sec:conclusions}.

\section{Preliminaries}\label{sec:prelim}

In our setup, we consider a finite-dimensional system prepared in a quantum state $\omega$ over which it is implemented a measurement defined by the Positive-Operator Valued Measure (POVM) $\{F_k\}_{k=1}^m$, a set of positive operators summing to the identity. The actual measurement represents an imperfect implementation of the target measurement, defined by the POVM $\{E_k\}_{k=1}^m$. Our objective is to control the distance between the ideal probabilities $\{ \mathrm{tr} (E_k \omega) \}_{k=1}^{m}$ and their experimentally estimated counterparts $\{\hat{f}_k\}_{k=1}^{m}$, with $\hat{f}_k \approx \mathrm{tr} (F_k \omega)$. To achieve this, we will use the Bretagnolle--Huber--Carol inequality \cite{wellner2013weak} and the notion of operational distance \cite{navascues2014how, puchala2018strategies, maciejewski2023exploring}.  

\subsection{Bretagnolle--Huber--Carol inequality}
The following concentration inequality provides an upper bound on the $l_1$ distance between a probability vector $\vec{p}$ associated with a multinomial distribution and its empirical estimate $\hat{f}$.  

\textbf{Bretagnolle--Huber--Carol inequality. } \textit{Let $\hat{Z} = (Z_1, Z_2, \ldots , Z_m)$ be a random vector distributed multinomially with parameters $\vec{p} = (p_1, p_2, \ldots , p_m)$ such that $\sum_{i=1}^m Z_i = N$. Then, with probability $1-\delta$ we have
\begin{align}
    || \hat{f} - \vec{p} ||_1 \leq \sqrt{ \frac{2}{N} \ln \frac{2^m}{\delta}}, \label{eq:bhc_bound}
\end{align}
where $\hat{f} = \hat{Z}/N$.}

This inequality will be used to control the shot noise in the estimates $\hat{f}_k$ of the probabilities $\mathrm{tr}(F_k \omega)$.

\subsection{Operational distance}
The operational distance between two POVMs is defined as the largest possible $l_1$ distance between the probability distributions they generate when measured on the same state.  

\textbf{Operational distance. } \textit{Let $\{E_k \}_{k=1}^m$ and $\{F_k \}_{k=1}^m$ be two $m$-outcome POVMs acting on a Hilbert space $\mathcal{H}$. The operational distance $d_{\mathrm{op}}$ is defined as
\begin{align}
    d_{\mathrm{op}} (E, F) = \max_{\rho \in \mathcal{D}(\mathcal{H})} \frac{1}{2} \sum_{k=1}^m  |p_k - q_k|, \label{eq:dop}
\end{align}
where $p_k = \mathrm{tr}(E_k \rho)$, $q_k = \mathrm{tr}(F_k \rho)$ and $\mathcal{D}(\mathcal{H})$ is the set of density matrices acting on $\mathcal{H}$. }

An equivalent expression is
\begin{align}\label{eq:dop_spectral_norm}
    d_{\mathrm{op}} (E, F) = \max_{x \in \mathcal{P}(m)} \left\Vert \sum_{k \in x} (E_k - F_k) \right\Vert,
\end{align}
where $\mathcal{P}(m)$ is the power set of $\{1, 2, ..., m\}$ and $\Vert \cdot \Vert$ denotes the spectral norm. We will use $d_{\mathrm{op}}$ to quantify the effect of measurement imperfections in the estimation of probabilities.  

\section{Results}\label{sec:results}

We now present a protocol to bound convex functions of quantum states that takes into account scenarios in which measurement devices are not perfectly controlled but operate with bounded inaccuracy, quantified by the operational distance $d_{\mathrm{op}}$.  

\textbf{Method.} \textit{Consider a density operator $\omega$ on a system of dimension $d$. Let $\{E_k\}_{k=1}^{m}$ define a POVM on the system and $\{ F_k \}_{k=1}^{m}$ any imperfect implementation such that $d_{\mathrm{op}}(E, F) \leq \epsilon_2 / 2$. Let $\hat{f}_k$ be the experimental estimation of $\mathrm{tr}(F_k \omega)$ obtained by measuring $\{ F_k \}_{k=1}^m$ across $N$ copies of $\omega$. Let $\mathcal{F}$ be a function of density matrices, and $\delta \in (0, 1]$. Then, the solution to the optimization problem
    \begin{align}
        \mathcal{F}_{\rm LB(UB)} = \min_{\rho } \; (\max_{\rho}) \; & \mathcal{F}(\rho) \nonumber\\
        \mathrm{s. t. } \quad &  \mathrm{tr}(\rho) = 1 \nonumber\\
        & \rho \geq 0  \nonumber\\
        & \sum_{k=1}^{m} |\mathrm{tr}(E_k \rho) - \hat{f}_k| \leq \epsilon_1 + \epsilon_2 \label{eq:l1_bound}
    \end{align}
is a lower (upper) bound for $\mathcal{F}(\omega)$ with probability at least $1-\delta$. Here 
\begin{align}
    \epsilon_1 = \sqrt{ \frac{2}{N} \ln \frac{2^m}{\delta}}.
\end{align}
}

The constraint on the probabilities follows from the triangle inequality together with Eqs.~\eqref{eq:bhc_bound}~and~\eqref{eq:dop},
\begin{align}
    \sum_{k=1}^{m} |\mathrm{tr}(E_k \rho) - \hat{f}_k| \leq & \sum_{k=1}^{m} |\mathrm{tr}(F_k \rho) - \hat{f}_k| \nonumber \\
    & +\sum_{k=1}^{m} |\mathrm{tr}(E_k \rho) - \mathrm{tr}(F_k \rho)|  \nonumber \\
    \leq & \sqrt{ \frac{2}{N} \ln \frac{2^m}{\delta}}  +  2 d_{\mathrm{op}}(E, F). 
\end{align}
This means that, given a quantum state $\rho$, the distance between the exact probabilities associated with $E$ and the estimated probabilities associated with $F$ is bounded by the finite-statistics error in estimating $\{\mathrm{tr}(F_k \omega)\}_{k=1}^m$ plus the distance between the POVMs $E$ and $F$.  

The constraints in the optimization problem define a convex set that contains the experimental quantum state with probability $1-\delta$, that is, a confidence region for the state. Minimizing or maximizing a function over this region yields a bound for the value of the function evaluated on the experimental quantum state $\omega$ with probability $1-\delta$. The larger the region, the more states are compatible with the constraints, and the looser the bounds. Therefore, the values of $\epsilon_1$ and $\epsilon_2$ should be as small as possible. However, care must be taken, since choosing values that are too small may result in an infeasible problem, where no quantum state satisfies the constraints.  

It is worth noting that, to apply the method to an arbitrary function, it is not necessary to measure an informationally complete (IC) POVM. This simplifies the experimental procedure, since IC-POVMs typically require an exponentially large number of measurement settings. Another advantage of the method is that, when the function $\mathcal{F}$ is convex or concave, the resulting optimization problem is itself convex and can be efficiently solved using interior-point methods \cite{boyd2004convex}.

To apply the method, we need to establish an upper bound for $d_{\mathrm{op}}(E, F)$. For POVMs with a small number of elements, we can use our knowledge of experimental noise sources to bound $d_{\mathrm{op}}(E, F)$ through numerical simulations. However, computing the analytical formula for $d_{\mathrm{op}}$ for POVMs with a large number of outcomes $m \gg 1$ may be challenging, as it involves a maximization over $\mathcal{P}(m)$, as shown in Eq.~\eqref{eq:dop}. For $n$-qubit systems, if the POVMs decompose as tensor products of local POVMs, $\{E_k \}_{k} = \{E_{k_1}^{(1)}\} \otimes ... \otimes \{E_{k_n}^{(n)}\}$ and $\{F_k \}_{k} = \{F_{k_1}^{(1)}\} \otimes ... \otimes \{F_{k_n}^{(n)}\}$ for $k=k_1 ... k_n$, we can bound $d_{\mathrm{op}}(E, F)$ as a sum of local operational distances \cite{puchala2018strategies, maciejewski2023exploring}:
\begin{align}
    d_{\mathrm{op}} \left(\bigotimes_{i=1}^n E^{(i)}, \bigotimes_{i=1}^n F^{(i)} \right) \leq \sum_{i=1}^n d_{\mathrm{op}} (E^{(i)}, F^{(i)}). \label{eq:localdop_bound}
\end{align}
This greatly simplifies the calculations, as it reduces the task to computing $d_{\mathrm{op}}$ for the smaller local POVMs separately.  

Quantum measurement tomography enables the reconstruction of an unknown POVM from experimental statistics \cite{fiuravsek2001maximum, d2004quantum, lundeen2009tomography, barbera2025boosting, zambrano2025fast}. In particular, Ref.~\cite{zambrano2025fast} shows that any $m$-outcome POVM acting on a $d$-dimensional system can be reconstructed with error at most $\epsilon$ in operational distance using $N \in \mathcal{O} \left(\frac{d^3 m}{\epsilon^2} \ln(d) \right)$ shots. Thus, a reconstruction $F$ of the POVM $E$ and a bound for $d_{\mathrm{op}}(E, F) \leq \epsilon$ can be obtained directly from this protocol.  

Other bounds follow from the relation
\begin{align}
    d_{\mathrm{op}}(E, F) &= \max_{\rho \in \mathcal{D}(\mathcal{H})} \frac{1}{2} \sum_{k=1}^m \left| \mathrm{tr}\big((E_k - F_k)\rho\big) \right| \nonumber \\
    &\leq \frac{1}{2} \sum_{k=1}^m \|E_k - F_k\|, \label{eq:opnorm_bound}
\end{align}
To obtain experimental bounds of Eq.~\eqref{eq:opnorm_bound}, we have two options. From Ref.~\cite{maciejewski2023exploring}, 
\begin{align}
   \frac{\sqrt{d(d+1)}}{2}\sum_{k=1}^m \sqrt{\mathbb{E}_{\psi} \!\left[ \langle \psi |E_k - F_k| \psi \rangle^2 \right]} = d\, d_{av},
\end{align}
where 
$d_{av}=\frac{1}{2d}\sum_{k=1}^m \sqrt{\|E_k - F_k\|_{F}^2 + \mathrm{tr}(E_k - F_k)^2}$, 
and the average is over states $|\psi\rangle$ forming a $2$-design. Since $\Vert \cdot  \Vert \leq \Vert \cdot  \Vert_F$, from  Eq.~\eqref{eq:opnorm_bound} we obtain
\begin{align}
    d_{\mathrm{op}}(E, F) &\leq \frac{1}{2}\sum_{k=1}^m \sqrt{\|E_k - F_k\|_{F}^2 + \mathrm{tr}(E_k - F_k)^2} \nonumber \\
    &= \frac{\sqrt{d(d+1)}}{2} \sum_{k=1}^m \sqrt{\mathbb{E}_{\psi}\langle \psi |E_k - F_k| \psi \rangle^2}. \label{eq:d_av_bound}
\end{align}
Thus, $d_{\mathrm{op}}(E, F)$ can be bounded by estimating the terms $\mathbb{E}_{\psi}\!\left[\langle \psi |E_k - F_k| \psi \rangle^2 \right]$. This requires preparing a set of states $\{|\psi_i\rangle\}$ forming a $2$-design and measuring them using $F$ to estimate $\{\langle \psi_i|F_k|\psi_i\rangle\}_{k=1}^m$. The corresponding values for $\{\langle \psi_i|E_k|\psi_i\rangle\}_{k=1}^m$ are obtained numerically, and the mean of the differences is then computed.  

Finally, another way to bound $d_{\mathrm{op}}(E, F)$ experimentally is to assume that the elements of the POVM $E$ are rank-one projectors, $E_k = a_k |\alpha_k \rangle \langle \alpha_k |$, with $a_k \geq 0$, and that the elements of the POVM $F$ are positive semidefinite with $\mathrm{tr}(F_k) = a_k$. This could be the case when $F_k = \Lambda (E_k)$ for some CPTP map $\Lambda$.  Then, from Eq.~\eqref{eq:opnorm_bound}, we have
\begin{align}
    d_{\mathrm{op}} (E, F) \, \leq & \,\frac{1}{2} \sum_{k=1}^m \sqrt{a_k^2 - a_k \langle \alpha_k | F_k | \alpha_k \rangle }. \label{eq:fid_bound}
\end{align}
The quantity $\langle \alpha_k | F_k | \alpha_k \rangle$ is a fidelity, which can be estimated by preparing the state $| \alpha_k \rangle$ and measuring it with the POVM $F$. 
%

From the results above, we see that there are several possible ways to obtain the $\epsilon_2$ term in our method, depending on the knowledge and control of the experimental setup. If the noise affecting the measurement device is well characterized, an upper bound for the error can be estimated numerically using Eq.~\eqref{eq:dop_spectral_norm} or \eqref{eq:opnorm_bound}, together with Eq.~\eqref{eq:localdop_bound}. If no information about the noise is available, quantum measurement tomography can be used to characterize it, or the discrepancy between the target and experimental POVMs can be measured directly using Eqs.~\eqref{eq:d_av_bound} or \eqref{eq:fid_bound}. If the bound for $d_{\mathrm{op}}$ is valid with probability $1-\delta$, then the bounds obtained with our method will hold with probability $1-2\delta$ instead of $1-\delta$. 

\section{Applications}\label{sec:numerical}

In this section, we illustrate our method with three numerical examples. The first concerns the certification of state preparation fidelity under noisy measurements. The second and third address the estimation of magnetization in a spin system and the certification of entanglement with a simple two-qubit witness, respectively, showing in both cases how neglecting measurement imperfections can lead to incorrect conclusions about the system.

\subsection{Minimum preparation fidelity}
\begin{figure} [t]                    
\centering
\includegraphics[width=.85\columnwidth, angle=0]{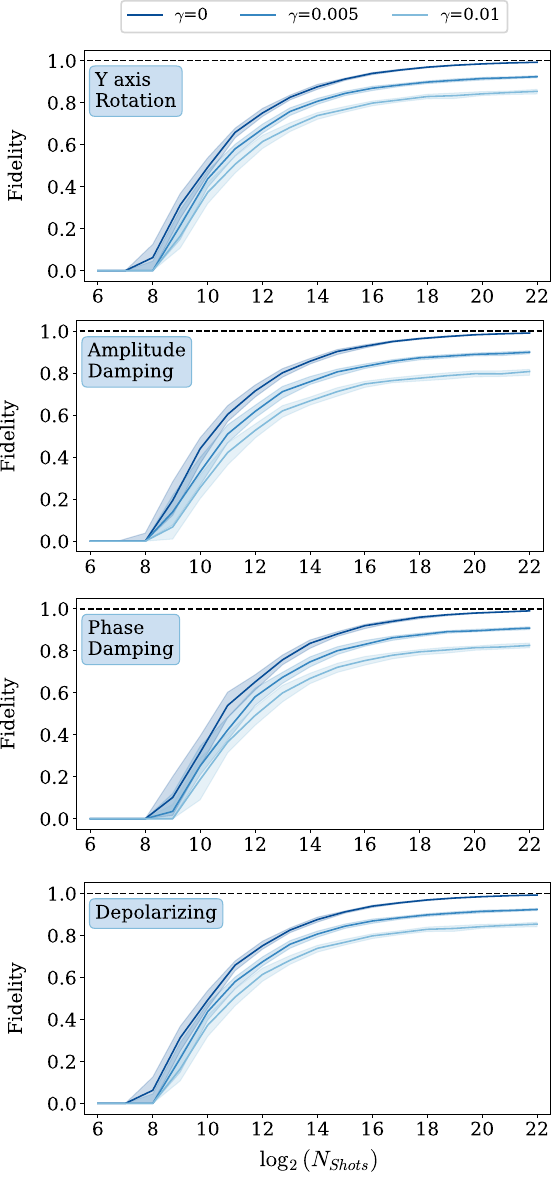}
\caption{Minimum two-qubit state fidelity as a function of the total number of shots $N_{shots}$, for a confidence level of $0.997$. From top to bottom: $Y$-axis rotation shift, amplitude damping, phase damping and depolarizing channels. The operational distance is bounded by Eq.~\eqref{eq:fid_bound}. Solid lines represent the median value obtained from 100 Haar-random pure states, while the shaded areas depict the interquartile ranges. Results are displayed for different values $\gamma$ of the corresponding channel (Appendix \ref{app:noise_model}).}
\label{fig:Channels_FId}
\end{figure} 

As a first numerical example, we assume that our goal is to experimentally prepare a pure quantum state $|\psi \rangle$. To check the quality of the preparation, we perform (imperfect) measurements on the system and then minimize the fidelity $\langle \psi | \rho | \psi \rangle$ between the states $\rho$ compatible with the measurements and the target state $| \psi \rangle$. This provides a certified lower bound for the fidelity between the target and experimental states.

We prepared $100$ Haar-random two-qubit pure states $|\psi \rangle$ and defined an ideal POVM constructed from tensor products of single-qubit SIC-POVMs. To implement noisy POVM measurements, we assume that the noise affecting the experiment is local. In this case, the ideal POVM elements $E_i$ are related to the noisy POVM elements $F_i$ through a quantum channel $\mathcal{N} =  \mathcal{N}^{(1)} \otimes \mathcal{N}^{(2)}$, that is, $F_i =  \mathcal{N}^{(1)} (E_i^{(1)}) \otimes \mathcal{N}^{(2)} (E_i^{(2)})$. In particular, we consider single-qubit amplitude damping, phase damping, depolarizing noise, and $Y$-axis rotation shifts, with the associated Kraus operators given in App.~\ref{app:noise_model}. For different noise intensities, we compute values for $\epsilon_2$ using Eq.~\eqref{eq:fid_bound} and simulate POVM measurements for different numbers of shots. We then apply Method \eqref{eq:l1_bound}, minimizing the fidelity:
\begin{align}
        F_{\min} = \min_{\rho \geq 0 } \quad & \langle \psi | \rho | \psi \rangle \nonumber\\
        \mathrm{s. t. } \quad &  \mathrm{tr}(\rho) = 1 \nonumber\\
        & \sum_{k=1}^{m} |\mathrm{tr}(E_k \rho) - \hat{f}_k| \leq \epsilon_1 + \epsilon_2.
\end{align}
The results are displayed in Fig.~\ref{fig:Channels_FId}. 

We observe that, for a sufficient number of shots, the method provides non-trivial lower bounds for the preparation fidelity. The quality of these bounds depends on the strength of the noise in the measurements, as expected.

\subsection{Magnetization under imperfect measurements}

\begin{figure}[t]
\centering
\includegraphics[width=1\columnwidth, angle=0]{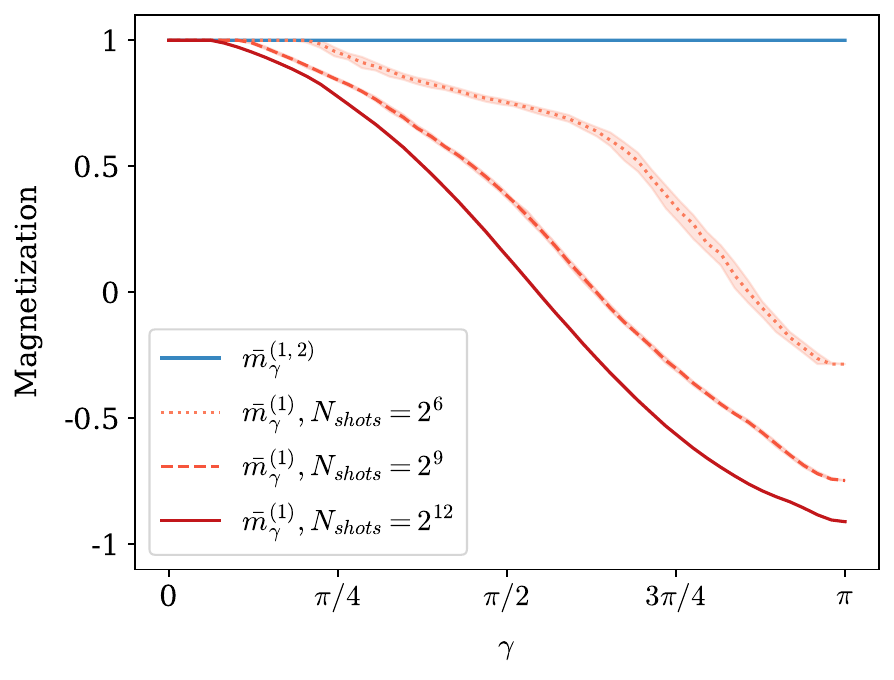}
\caption{Maximum spin magnetization as a function of the noise rotation angle $\gamma$, for $5$ qubits and a confidence level of $0.997$. Solid lines represent the median value obtained from 100 Haar-random pure states, while the shaded areas depict the interquartile ranges. The results that account for measurement noise are shown in blue, while the red curves correspond to results that ignore operator noise, with varying $N_{shots}$.}
\label{fig:Magnetization}
\end{figure}

Here we show that neglecting measurement imperfections in an experiment can lead to incorrect conclusions about the system. 

Consider a quantum ferromagnetic system prepared in the state $\omega=\otimes_{i=1}^n \ket{0}\bra{0}_i$. The magnetization operator is defined as 
\begin{equation}
M=\frac{1}{n}\sum_{i=1}^n \sigma^z_i\,,
\end{equation} 
where the Pauli matrix $\sigma_i^z$ acts only on the $i$th spin. The magnetization is then 
\begin{align}
\bar{m}={\rm tr}(M \omega)=1.
\end{align}

We want to determine whether the system is fully magnetized or not, based on finite statistics and imperfect measurements. Let us consider the set of target POVM elements
\begin{eqnarray}
\{E_k\}_{k=1}^{2^n}=\{\ket{z_1}\bra{z_1}\otimes \ket{z_2}\bra{z_2} \otimes \dots \otimes \ket{z_n}\bra{z_n}\}_{z_i \in \{1,0\}}
\nonumber
\end{eqnarray}
and assume that the actual experimentally implemented POVM elements are rotated versions of the target ones around the $y$ axis by an angle $\gamma \in [0, \pi]$, i.e.,
\begin{equation}
F_k={U_y}^\dag E_k U_y\,,
\end{equation} 
with $U_y=\otimes_{i=1}^n e^{-i \sigma^y_i \gamma}$.
This corresponds to an unwanted rotation of the measurement setup.

The optimization problem that calculates the maximum possible magnetization compatible with the experiment is
\begin{align}
        \bar{m}_{\gamma} = \max_{\rho \geq 0 } \quad & {\rm tr}(M \rho) \nonumber\\
        \mathrm{s. t. } \quad &  \mathrm{tr}(\rho) = 1 \nonumber\\
        & \sum_{k=1}^{m} |\mathrm{tr}(E_k \rho) - \hat{f}_k| \leq \epsilon \label{eq:m-syst}
\end{align}
with $\hat{f}_k$ the experimental estimate of $\mathrm{tr}(F_k \omega)$ and $\epsilon$ a bound for the noise. Taking into account both systematic and statistical errors, we have $\epsilon= \epsilon_1 + \epsilon_2$, with $\epsilon_1$ the bound for shot noise and $\epsilon_2$ the bound for imperfect measurements. We denote the solution to this problem as $\bar{m}_{\gamma}^{(1,2)}$. Instead, if we consider only statistical errors we have $\epsilon = \epsilon_1$, that is, $\epsilon_2 = 0$,  and we denote the solution as $\bar{m}^{(1)}_{\gamma}$. 

In Fig.~\ref{fig:Magnetization}, fixing $1-\delta=0.997$, we show the behavior of the maximum magnetizations $\bar{m}_{\gamma}^{(1,2)}$ and $\bar{m}_{\gamma}^{(1)}$ as functions of the rotation angle $\gamma$ for a system of 5 spins. The bound $\epsilon_2$ depends on $\gamma$ and is calculated using Eq.~\eqref{eq:fid_bound}. We observe that while $\bar{m}^{(1,2)}_{\gamma} = 1$ for all $\gamma$, $\bar{m}_{\gamma}^{(1)}$ rapidly decreases below one as $\gamma$ increases. Hence, ignoring measurement imperfections may lead to strongly misleading conclusions about the state, since one could be highly confident that the state has magnetization far from one, or even negative, a conclusion that is in fact false. As a remark, $\bar{m}^{(1,2)}_{\gamma}$ remains unchanged for different numbers of shots, so we only display the case $N_{shots}=2^{12}$.

\subsection{Entanglement witness under imperfect measurements}
\begin{figure}[t]
\centering
\includegraphics[width=0.99\columnwidth, angle=0]{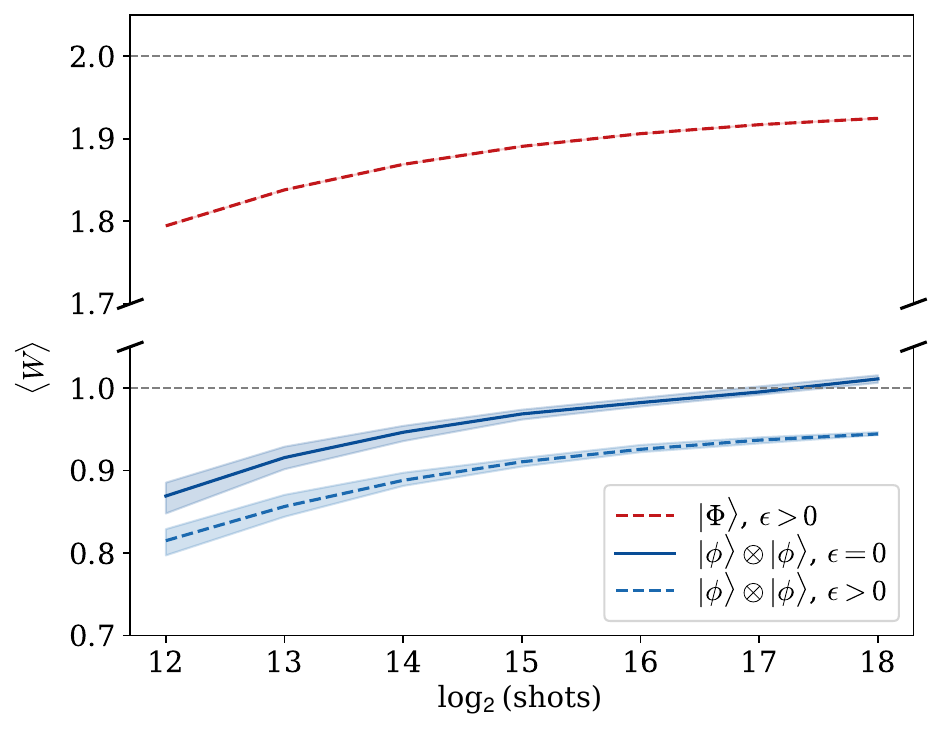}
\caption{Minimum value of the witness as a function of the total number of shots $N_{shots}$, for $2$ qubits and a confidence level of $0.997$. From top to bottom: entangled state considering measurement noise, separable state ignoring noise, and separable state considering measurement noise. The operational distance is bounded by Eq.~\eqref{eq:fid_bound}. Solid lines represent the median value obtained from 100 Haar-random pure states, while the shaded areas depict the interquartile ranges.}
\label{fig:bell_inequality}
\end{figure}

We now consider the task of certifying entanglement using a two-qubit witness, and show that neglecting systematic errors can lead to falsely identifying separable states as entangled.

Consider the entanglement witness 
\begin{align}
    W = X_1 \otimes X_2 + Z_1 \otimes Z_2.
\end{align}
For separable states we have $W \leq 1$, while for entangled states $W \leq 2$. 
Thus, if for a state $\omega$ we find $\mathrm{tr}(\omega W) > 1$, we can certify that the state is entangled~\cite{morelli2022entanglement}. 

The target POVM $\{ E_k \}_{k=1}^{8}$ is defined by the (separable) eigenstates of $X \otimes X$ and $Z \otimes Z$.  We introduce imperfect observables following Ref.~\cite{morelli2022entanglement}:
\begin{align}\label{eq:imperfect_measurements_witness}
X_1^{(\mathrm{imp})} = X_2^{(\mathrm{imp})} &= \cos (\theta_x) X + \sin (\theta_x) Z, \nonumber \\
Z_1^{(\mathrm{imp})} = Z_2^{(\mathrm{imp})} &= \cos (\theta_z) X + \sin (\theta_z) Z.
\end{align}
This defines an imperfect POVM $\{F_k\}_{k=1}^{8}$ consisting of the eigenstates of 
$X_1^{(\mathrm{imp})} \otimes X_2^{(\mathrm{imp})}$ and 
$Z_1^{(\mathrm{imp})} \otimes Z_2^{(\mathrm{imp})}$. We measure the state $\omega$ of the system using this POVM with a total of $N$ shots, obtaining frequencies $\hat{f}$. The following optimization problem then provides a certified lower bound for $\mathrm{tr}(\omega W)$:
\begin{align}
        \min_{\rho \geq 0} \quad & {\rm tr}(W \rho) \nonumber\\
        \mathrm{s.t.} \quad &  \mathrm{tr}(\rho) = 1 \nonumber\\
        & \sum_{k=1}^{m} \big|{\rm tr}(E_k \rho) - \hat{f}_k\big| \leq \epsilon, 
\end{align}
where $\epsilon = \epsilon_1 + \epsilon_2$, with $\epsilon_1$ the bound for shot noise and $\epsilon_2$ the bound for imperfect measurements.  

If ${\rm tr}(W \rho) > 1$, the state is entangled. However, with imperfect measurements it is possible to wrongly identify separable states as entangled. 
This effect is illustrated in Fig.~\ref{fig:bell_inequality}, where we consider imperfect observables given in Eq.~\eqref{eq:imperfect_measurements_witness} with $\theta_x = 0.01$, $\theta_z = \pi/2 - 0.01$, and the separable state $|\phi \rangle \otimes |\phi \rangle$, with $|\phi\rangle = \cos \tfrac{\pi}{8}|0\rangle + \sin \tfrac{\pi}{8}|1\rangle$.

For $N = 2^{18}$, ignoring experimental errors ($\epsilon_2 = 0$) incorrectly certifies entanglement, while the considering measurement imperfections ($\epsilon_2 > 0$) does not.  
We also observe that for the maximally entangled state $|\Phi^{+}\rangle = \tfrac{1}{\sqrt{2}}(|00\rangle + |11\rangle)$, the corrected method continues to provide meaningful results, certifying entanglement. The bound $\epsilon_2$ depends on $\theta_x$ and $\theta_z$ and is calculated using Eq.~\eqref{eq:fid_bound}.

When either the experimental noise or the number of shots becomes large, the method that neglects measurement imperfections ($\epsilon_2 = 0$) breaks down, since the corresponding SDP becomes infeasible and no bounds can be derived. Our proposal, which accounts for imperfections through the operational norm $d_\mathrm{op}$, remains valid in all regimes and always yields certified bounds.

\section{Conclusions}\label{sec:conclusions}

We have introduced a certification method for convex functions of quantum states that addresses both shot noise and measurement imperfections in quantum systems. Our method extends the confidence region for quantum tomography introduced in Ref.~\cite{zambrano2024certification} to situations where the control of the experiment is not perfect, but inaccuracies in operations are bounded. This addition prevents wrong conclusions about the experiment, as we showed in our numerical examples, and ensures that certified results can always be obtained even in regimes where previous methods fail.

The parameters defining the confidence region can be obtained through accessible means, either by numerical simulation of the experiment or by employing any of the experimental alternatives proposed. The method does not require the use of a specific POVM, allowing experimentalists to choose measurements freely, aiming for simplicity depending on the experimental setup. Moreover, the measurements required by the method do not need to be informationally complete, further reducing the experimental complexity of the protocol. Finally, the optimization problem required by the method can be efficiently solved using convex optimization techniques.

Compared with existing approaches, which either neglect systematic errors or restrict to specific noise models, our framework jointly incorporates both statistical and systematic imperfections in a fully general way. While we focused on fidelities, entanglement witnesses, and magnetization as examples, the method applies to a broad class of convex quantities of interest, such as energies, entropies, or quantum Fisher information. An interesting direction for future work is to explore extensions to non-convex functions through convex relaxations.

Overall, our proposal provides a versatile and robust certification tool that is directly applicable to near-term quantum experiments. In particular, it offers a practical way to obtain rigorous guarantees in realistic settings where devices are noisy and calibration is limited, which is essential for the reliable development of quantum technologies.

\begin{acknowledgments}
This work was supported by the Government of Spain (Severo Ochoa CEX2019-000910-S, FUNQIP, and European Union NextGenerationEU PRTR-C17.I1), Fundació Cellex, Fundació Mir-Puig, Generalitat de Catalunya (CERCA program), the EU Quantera project Veriqtas, the AXA Chair in Quantum Information Science, the ERC AdG CERQUTE and the EU and Spanish AEI project QEC4QEA.
\end{acknowledgments}

\bibliography{bib}

\appendix

\section{Kraus operators for the noise models \label{app:noise_model}}

Y-axis rotation shift:
\begin{align}
    K_0 &= \begin{bmatrix}
        \cos(\gamma/2) & -\sin(\gamma/2) \\
        \sin(\gamma/2) & \cos(\gamma/2)
    \end{bmatrix}
\end{align}

Amplitude damping:
\begin{align}
    K_0 &= \begin{bmatrix}
        1 & 0 \\
        0 & \sqrt{1 - \gamma}
    \end{bmatrix}, \\
    K_1 &= \begin{bmatrix}
        0 & \sqrt{\gamma} \\
        0 & 0
    \end{bmatrix}.
\end{align}

Phase damping:
\begin{align}
    K_0 &= \sqrt{1 - \gamma} \, I, \\
    K_1 &= \begin{bmatrix}
        \sqrt{\gamma} & 0 \\
        0 & 0
    \end{bmatrix}, \\
    K_2 &= \begin{bmatrix}
        0 & 0 \\
        0 & \sqrt{\gamma}
    \end{bmatrix}.
\end{align}

Depolarizing:
\begin{align}
    K_0 &= \sqrt{1 - \gamma} \, I, \\
    K_1 &= \sqrt{\frac{\gamma}{3}} \, X, \\
    K_2 &= \sqrt{\frac{\gamma}{3}} \, Y, \\
    K_3 &= \sqrt{\frac{\gamma}{3}} \, Z,
\end{align}

\end{document}